\documentstyle[12pt]{article}
\newcommand{\beq}{\begin{equation}}
\newcommand{\eeq}{\end{equation}}
\newcommand{\bdm}{\begin{displaymath}}
\newcommand{\edm}{\end{displaymath}}
\newcommand{\bea}{\begin{eqnarray}}
\newcommand{\eea}{\end{eqnarray}}

 \newcommand{\reset}{  \setcounter{equation}{0}  }
\newcommand{\th}{\theta}

\newcommand{\ib}{\bar{\imath}}
\newcommand{\jb}{\bar{\jmath}}

 \newcommand{\NP}{{\em Nucl. Phys. }}
 
\begin{document}
\setcounter{page}{0}
\topmargin 0pt
\oddsidemargin 5mm
\renewcommand{\thefootnote}{\fnsymbol{footnote}}
\newpage
\setcounter{page}{0}
\begin{titlepage}
\begin{flushright}
Swansea SWAT/93-94/28 \\
USP-IFQSC/TH/94-03\\
hep-th/9404188
\end{flushright}
\vspace{0.5cm}
\begin{center}
{\large {\bf Boundary Bound States in Affine Toda Field Theory}} \\
\vspace{1.8cm}
{\large Andreas Fring
\footnote{e-mail address: A.Fring@UK.AC.SWANSEA}   and
 Roland K\"oberle\footnote{ Supported in part by CNPq-Brasil.  e-mail address:
ROLAND@IFQSC.USP.SC.BR } } \\
\vspace{0.5cm}
{\em $  ^{*} $ Department of Physics,
University College of Swansea,
Swansea SA2 8PP,UK} \\
{\em $ ^{\dag}$ Universidade de S\~ao Paulo,
Caixa Postal 369, CEP 13560 S\~ao Carlos-SP, Brasil\\}
\vspace{3cm}
\renewcommand{\thefootnote}{\arabic{footnote}}
\setcounter{footnote}{0}
\begin{abstract}
{ We demonstrate that the generalization of the  Coleman-Thun mechanism
may be applied to the situation, when considering scattering processes in
1+1-dimensions in the presence of reflecting boundaries.  For affine Toda
field theories we  find that the binding energies of the bound states are
always half the sum over a set of  masses having the same colour with respect
to the bicolouration of the Dynkin diagram. For  the case of  $E_6$-affine Toda
field  theory we compute explicitly the spectrum of all higher boundary bound
states. The complete set of  states constitutes a closed  bootstrap. }
\end{abstract}
\vspace{.3cm}
\centerline{April 1994}
 \end{center}
\end{titlepage}
\newpage
\section{Introduction}

In any quantum field theory its general principles will impose restrictions on
their possible scattering matrices. It has turned out that in an infinite two
dimensional space-time volume in particular, the equations resulting from
integrability, like the Yang-Baxter- \cite{YB} and bootstrap equations
\cite{ZZ}, serve as a very  powerful tool and lead to their exact
determination.
Various  questions of physical   interest,  like the study of dissipative
quantum
mechanical systems \cite{Leg}, the study of  the space of  boundary states in
open string theory \cite{Callan,Witten} and that of several statistical systems
require the  restriction to a finite
space dimension. Then in addition to the S-matrix a matrix, say W, will enter
the formalisms encoding
the scattering off the boundary.
For many theories the approach, which has been persued for infinite volume,
has been
successful in the finite volume case, where one solves the
analogous set of equations. This yields an explicit expression
for the W-matrices [6-17].  For
the Sine-Gordon theory the results have been confirmed by the Bethe Ansatz
\cite{FS}.
\par
In this paper we are concerned  with their generalizations,
affine Toda field theories \cite{MOP}  for which it was demonstrated in
\cite{FK,FK1},
that it is possible to determine the W-matrices by employing the
crossing, unitarity and bootstrap equations.
The methods of solving the bootstrap equation in \cite{FK} was rather indirect
using Fourier
transforms and in \cite{FK1}  more explicit use has been made of the crossing
unitarity relation
\cite{GZ}. But hitherto no sufficient explanation has been provided concerning
the
interpretation of the
singularities occuring in the W-matrix.  The natural conjecture is to assume
that we may employ a bootstrap program in analogy to the bulk theory.  This
means
that ``bootstrapping" on odd order poles  with positive residues should yield
other matrices,
such that all further matrices obtained in this manner form a closed system in
the sense specified below.  This prescription led to a generalization
\cite{Muss,BCDS}
of the Coleman-Thun mechanism \cite{Coleman} in the bulk theory and provided a
complete interpretation of the multipole structure in the S-matrix.
The principal aim of the following is to show that in the theory with
boundaries
the
same program may be applied.
\par
Our presentation is organised as follows. In section 2 we briefly review some
of
the main features of the S-and W-matrices. Section 3 is devoted to a general
discussion of the boundary bound state bootstrap equation, from which we
compute
the binding energies of all boundary bound states.
 We illustrate  the general arguments in
section 4 with the concrete example of the $E_6$-affine Toda field theory.
Here we verify that the bootstrap
program
may be applied to the theory in the presence of reflecting boundaries in the
same
fashion as in the bulk theory, which is our main result. In section 5 we state
our conclusions.
\par
\section{Preliminaries}
\reset
In order to establish our notation and to achieve a somewhat self-consistent
presentation we shall briefly review some of the established properties of the
S- and W-matrices, which may be derived from the Zamolodchikov algebra
\cite{ZZ}.
The W-matrix will always carry two indices, i.e. $W_{i \alpha} (\theta_i) $,
the
first  refering to particle of type $i$ and the second indicates that the
boundary is
in state $\alpha$. For additional distinction we shall always be refering to
particles and  boundary
bound states by latin and greek letters, respectively.
Simply by applying twice the relations for the Zamolodchikov algebra
one obtains  the unitarity conditions
\beq
S_{ij}(\th) S_{ij}(-\th) \; =\; 1 \qquad \hbox{and} \qquad W_{i\alpha}(\th)
W_{i\alpha}(-\th)
\;= \;1 \; .
\eeq
Far less obvious are the crossing relations for the S- \cite{ZZ} and W-matrices
\cite{GZ}
\beq
S_{i \jb}(\th) \;= \;S_{ij} ( i \pi - \th) \qquad \hbox{and} \qquad  W_{i
\alpha}( \th)
 W_{\ib \alpha} (\th +i \pi) \; =\;  S_{ii} (2 \th) \; . \label{eq: cross}
\eeq
The most powerful restrictions for diagonal scattering matrices result from
the so-called bootstrap equations. For  the S-matrix  they read \cite{ZZ}
\beq
S_{lk} ( \th ) \; = \;  S_{li}\left( \th + i \eta_{ik}^{j} \right) \;
S_{lj}\left( \th - i \eta_{jk}^{i}
\right) \;    \label{eq: boots}
\eeq
and for the W-matrix  \cite{FK}
\beq
W_{k \alpha} ( \th ) \; = \;  W_{i \alpha}\left( \th + i \eta_{ik}^{j} \right)
\; W_{j \alpha}
\left( \th - i \eta_{jk}^{i} \right)   S_{ij} (2 \th + i \eta_{ik}^{j} - i
\eta_{jk}^{i}    )
\;  \;\; .\label{eq: wboots}
\eeq
Here $\eta_{ij}^{k}$ denotes the fusing angle which emerges whenever the
particle $k$ can exist as a
bound state in the scattering between $i$ and $j$.
\par
The solutions to this equations contain several ambiguities, like the
CDD-ambiguity
known from the bulk theory \cite{CDD},  which in the context of affine Toda
field
theory corresponds to a shift of the rapidity
by  $i \pi$ \cite{FK1}. Furthermore, it was pointed out by Sasaki \cite{Sasaki}
that whenever we have a solution  to (\ref{eq: wboots}), say  $W_{n
\alpha} ( \th )$
for $n =1, \ldots ,r$,  ($r$ denoting the number of different types of
particles) then, because of
(\ref{eq: boots}),  $W_{n \alpha} ( \th )   \prod S_{n l} ( \th ) $
will be a solution as well.  However, the new solution might introduce
additional
poles in
the physical sheet, which have to be given an interpretation. It is the

\section{Boundary bound state bootstrap equation }
\reset

which will account for these \cite{GZ},
\beq
W_{j \beta} ( \th ) \; = \;  S_{ij} \left( \th + i \eta_{i \alpha}^{\beta}
\right)
 W_{j \alpha} ( \th ) S_{ij}  \left( \th - i \eta_{i \alpha}^{\beta}  \right)
\;\; . \label{eq: bbsb}
\eeq
Here $ \eta_{i \alpha}^{\beta}$ denotes the pole which emerges when particle
$i$
hits
the boundary in the state $\alpha$ and causes it to transit into state $\beta$.
The interpretation of this equation is the usual one and assumes
an analytic continuation in the rapidity plane. It says that the
 following two situations are equivalent: when $i$ possesses
precisely  this rapidity, an additional particle bouncing off the
wall may scatter off particle i either before or after the excitation
 $i  + \alpha \rightarrow \beta$ takes place.
In the latter case  it will pick up two
S-matrices. It is this equation we shall be mainly concerned with  in the
following. We depict  it in figure 1.
\par
Notice that the ambiguities in $W_{n\alpha} (\th)$ are restricted by this
equations. At first
we note that the factor $S_{nl}(\th)$ has to be the same for all possible
values
of $\alpha$,
but more severe restrictions result from the possible additional poles
introduced by it.
We shall comment on this point more below.
\par
\begin{figure}[h]
\setlength{\unitlength}{0.0125in}
\begin{picture}(40,140)(60,470)
\thicklines
\put(120,510){\line( 1,0){170}}
\put(340,510){\line(1,0){170}}
\put(120,500){\line(2,1){20}}
\put(150,500){\line(2,1){20}}
\put(160,500){\line(2,1){20}}
\put(170,500){\line(2,1){20}}
\put(180,500){\line(2,1){20}}
\put(190,500){\line(2,1){20}}
\put(220,500){\line(2,1){20}}
\put(230,500){\line(2,1){20}}
\put(240,500){\line(2,1){20}}
\put(250,500){\line(2,1){20}}
\put(260,500){\line(2,1){20}}
\put(270,500){\line(2,1){20}}
\put(340,500){\line(2,1){20}}
\put(350,500){\line(2,1){20}}
\put(380,500){\line(2,1){20}}
\put(390,500){\line(2,1){20}}
\put(400,500){\line(2,1){20}}
\put(410,500){\line(2,1){20}}
\put(420,500){\line(2,1){20}}
\put(430,500){\line(2,1){20}}
\put(440,500){\line(2,1){20}}
\put(450,500){\line(2,1){20}}
\put(480,500){\line(2,1){20}}
\put(490,500){\line(2,1){20}}
\put(310,540){$=$}
\put(220,510){\vector(1,3){25}}
\put(220,510){\line(-1,3){25}}
\put(410,510){\vector(1,3){25}}
\put(410,510){\line(-1,3){25}}
\put(170,510){\line(-2,1){50}}
\put(120,535){\vector(2,-1){20}}
\put(460,510){\line(-2,1){120}}
\put(340,570){\vector(2,-1){25}}
\put(125,535){$i$}
\put(120,516){$\eta_{i \alpha}^{\beta}$}
\put(187,575){$j$}
\put(390,575){$j$}
\put(350,568){$i$}
\put(138,501){$\alpha $}
\put(210,500){$\beta $}
\put(368,501){$\alpha $}
\put(470,500){$\beta $}
\end{picture}
 \caption{The boundary bound state bootstrap equation}
 \end{figure}
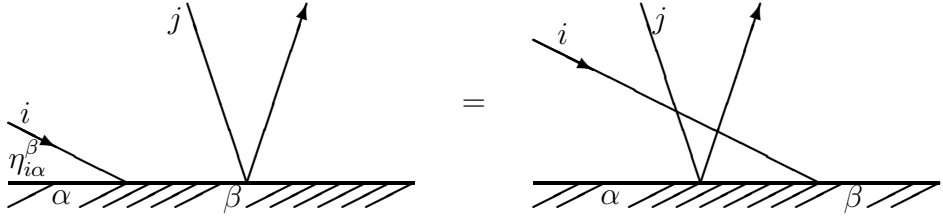
\par
We now like to obtain some information about the  possible angles $\eta$.
We apply equation (\ref{eq: bbsb}) to the two processes
$i + \alpha \rightarrow \beta$ and $\beta + \ib \rightarrow \alpha$
and use unitarity and crossing of the S-matrix to obtain
\beq
\eta_{i \alpha}^{\beta} + \eta_{ \ib \beta}^{\alpha} \; = \; \pi \;\;  .
\label{eq: zwei}
\eeq
A further useful equation results when considering a situation in which the
state $\beta$
results as an excitation from $\alpha$ by particle $i$ and $\gamma$ can be
reached from
$\alpha$ by $j$ and from $\beta$ by $k$, refer figure 2.
\begin{figure}[h]
\setlength{\unitlength}{0.0125in}
\begin{picture}(40,130)(60,500)
\thicklines
\put(150,510){\line(1,0){300}}
\put(150,570){\line(1,0){300}}
\put(150,610){\line(1,0){300}}
\put(140,508){$\alpha $}
\put(140,568){$\beta $}
\put(140,610){$\gamma $}
\put(205,540){$j $}
\put(305,590){$k $}
\put(405,540){$i $}
\thinlines
\put(200,510){\vector(0,1){100}}
\put(300,570){\vector(0,1){40}}
\put(400,510){\vector(0,1){60}}
\end{picture}
 \caption{Boundary transitions}
\end{figure}
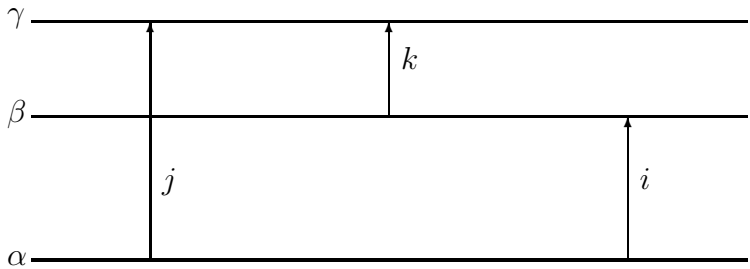
In this case we may
eliminate the
W-matrices in  (\ref{eq: bbsb}) and obtain an equation involving solely
S-matrices
$$
S_{lj} \left( \th  +  i \eta_{j \alpha}^{\gamma} \right)
S_{lj} \left( \th  \!- \!i \eta_{j \alpha}^{\gamma} \right)  \! =  \!
S_{li} \left( \th  +  i \eta_{i \alpha}^{\beta} \right) S_{li} \left( \th  \!-
\! i \eta_{i \alpha}^{\beta} \right)
S_{lk} \left( \th  +  i \eta_{k \beta}^{\gamma} \right) S_{lk} \left( \th
\!-\!
 i  \eta_{k \beta}^{\gamma} \right) . $$
This equation is useful since in general we consider the situation involving
the
boundaries,
when the bulk theory described by the S-matrices is already known.
 We shall employ  this equation
below as a consistency equation for the fusing angles. Clearly, similar
equations may
be derived involving other processes.
\par
We shall now compute the binding energies for the boundary bound states.
Remembering \cite{FK1} that the W-matrix is constituted out of blocks
${\cal{W}}_x$
\beq
{\cal W}_x ( \theta ) \; = \; \frac{ w_{1-x} ( \theta)  w_{-1-x}(\theta) } {
w_{1-x - B} ( \theta)
w_{-1-x + B} ( \theta) } \;\; ,\label{eq: subblock}
\eeq
with
\beq
w_x(\th)  = \frac{ < \frac{x-h}{2} >_{\th} } { < \frac{x-h}{2} >_{-\th} } \;
\eeq
and $< x>_{\th} = \sinh \frac{1}{2} \left( \th + \frac{i \pi}{h} x \right) $,
$
0 \leq B(\beta) \leq 2$. $  {\cal W}_x(\th)$
posses poles at  $\th_{\pm} = (\pm 1 - x - h)/ \frac{i \pi}{2h} $,
and we therefore obtain that a pole in the physical
sheet can only result when $x$ lies between $2h$ and $4h$.  ($h$ denotes here
as
usual the
Coxeter number of the underlying Lie algebra.) Recalling further that the
W-matrix was constructed
by replacing the blocks  $\{ x\}_{\th} = [ x ]_{\th}/ [x]_{-\th}$
($[x]_{\th} = < x+1>_{\th} <x-1>_{\th} / < x+1-B>_{\th} <x-1+B>_{\th} $) in
$S_{i \ib }$ by
${\cal{W}}_x(\th)$ and shifting some of the $x$'s by $2h$, we obtain that the
pole $\th_{\pm}  $
has a corresponding pole in $S_{ii}(\th)$ at $\eta_{\pm} = (\pm 1 + x ) \frac{i
\pi}{2h} $, which relates to one
in $W_{i \alpha}(\th) $ at $2 \eta_{\pm}$. We may see this more directly from
(\ref{eq: cross}).
\par
In other words, if the block ${\cal W}_{(x-h)+2h}(\th)$ occurs
in $W(\th)$, then the block $\{x\}_{\th} $ occurs in $S_{ii}(\th)$
and $\{h-x\}_\th$ appears in $S_{p\ib}(\th)$.
Having found a relation between the poles in the W- and in the
scattering matrix, we have shifted the problem of finding the angles
$\eta_{i \alpha}^{\beta}$ towards a problem of analysing the singularities in
$S_{ii}$ at $ 2 \eta_{\pm}$.
This means that we may view the two scattering matrices in (\ref{eq: bbsb}) as
part of a bootstrap equation. In fact we shall verify
\beq
 S_{li} \left( \th + i \eta_{i \alpha}^{\beta} \right) S_{li}  \left( \th - i
\eta_{i \alpha}^{\beta}  \right)
\; = \; \prod_k S_{lk} (\th)  \;\;\; , \label{eq:  Sbox}
\eeq
where all the k's will be of the same colour with respect to the bicolouration
of the Dynkin
diagram \cite{PD,FLO,FO}. We can understand this equation by drawing the
possible
graphs,
which are in agreement with the generalization of the Coleman-Thun mechanism
\cite{Coleman}
in the bulk theory.
Considering a situation in which two identical particles with relative
rapidities $2 \eta$
scatter against each other, we observe that the only consistent graphs we may
draw are
box-diagrams. Then the left hand side of this equation simply corresponds to
the
scattering
of an additional particle $l$ with the two particles and the right hand side
encodes the scattering
with the intermediate particles. We illustrate this in some more detail for a
double pole. The
only graph we may draw is depicted in figure 3.
\begin{figure}[h]
\setlength{\unitlength}{0.0125in}
\begin{picture}(40,140)(60,470)
\thicklines
\put(160,520){\line(1,0){60}}
\put(140,580){\line(1,0){100}}
\put(140,580){\line(1,-3){20}}
\put(240,580){\line(-1,-3){20}}
\put(160,520){\line(-2,-1){70}}
\put(140,580){\line(-2,1){50}}
\put(240,580){\line(2,1){50}}
\put(220,520){\line(2,-1){70}}
\put(90,605){\vector(2,-1){25}}
\put(240,580){\vector(2,1){25}}
\put(220,520){\vector(2,-1){35}}
\put(110,495){\vector(2,1){25}}
\put(160,520){\vector(1,0){30}}
\put(140,580){\vector(1,0){50}}
\put(140,580){\vector(1,-3){10}}
\put(240,580){\vector(-1,-3){10}}
\put(182,508){ $b$}
\put(182,588){ $a$}
\put(110,610){\vector(1,-4){30}}
\put(150,548){ $k$}
\put(233,548){ $\bar{k}$}
\put(110,548){ $l$}
\put(90,587){ $i$}
\put(90,495){ $i$}
\put(280,590){ $i$}
\put(280,495){ $i$}
\put(340,587){ $i$}
\put(340,495){ $i$}
\put(410,520){\line(1,0){60}}
\put(390,580){\line(1,0){100}}
\put(390,580){\line(1,-3){20}}
\put(490,580){\line(-1,-3){20}}
\put(410,520){\line(-2,-1){70}}
\put(390,580){\line(-2,1){50}}
\put(490,580){\line(2,1){50}}
\put(470,520){\line(2,-1){70}}
\put(340,605){\vector(2,-1){25}}
\put(490,580){\vector(2,1){25}}
\put(470,520){\vector(2,-1){35}}
\put(360,495){\vector(2,1){25}}
\put(410,520){\vector(1,0){30}}
\put(390,580){\vector(1,0){50}}
\put(390,580){\vector(1,-3){10}}
\put(490,580){\vector(-1,-3){10}}
\put(432,508){ $b$}
\put(434,588){ $a$}
\put(430,610){\vector(1,-4){30}}
\put(400,548){ $k$}
\put(483,548){ $\bar{k}$}
\put(430,548){ $l$}
\put(530,590){ $i$}
\put(530,495){ $i$}
\put(310,540){$=$}
\thinlines
\put(220,520){\line(1,0){40}}
\put(245,512){ $\eta$}
\put(160,520){\line(1,-3){8}}
\put(164,507){ $\Omega$}
\end{picture}
\caption{``Double pole bootstrap equation "}
\end{figure}
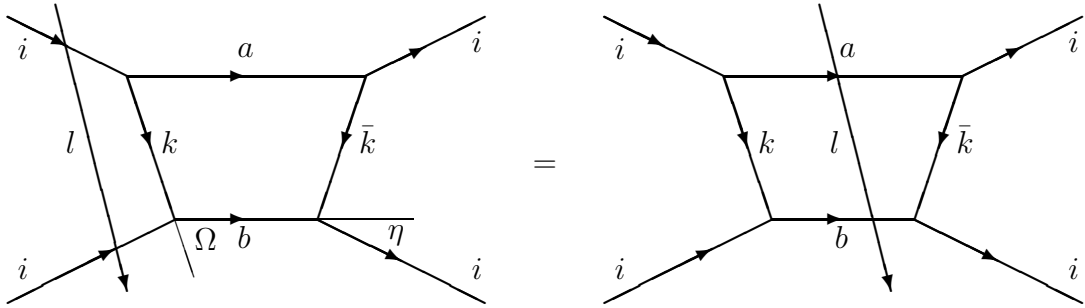
\par
It corresponds to the equation
\beq
S_{li} \left( \th + i \eta   \right) S_{li} \left( \th - i \eta   \right)  \; =
\; S_{la} \left( \th  \right)
S_{lb} \left( \th  \right)  \;\; .  \label{eq: sdouble}
\eeq
On the other hand we could have scattered $l$ with different rapidities such
that we obtain
the usual bootstrap equation involving three S-matrices only
\bea
 S_{li} \left( \th + i \eta \right) &=&S_{la} \left( \th  \right)S_{lk} \left(
\th + i \Omega \right) \\
S_{li} \left( \th - i \eta  \right)&=&S_{lb} \left( \th  \right) S_{l\bar{k}}
\left(i \pi + \th + i \Omega \right) \; ,
\eea
which when multiplied with each other give equation (\ref{eq: sdouble}). In the
usual fashion
\cite{FO} there are two fusing rules  associated with them $ \sigma^{\xi(i)}
\gamma_i =
\sigma^{\xi(a)} \gamma_a + \sigma^{\xi(k)} \gamma_k$ and  $
\sigma^{\tilde{\xi}(i)} \gamma_i
= \sigma^{\xi(b)} \gamma_b + \sigma^{ \tilde{\xi}(\bar{k})} \gamma_{\bar{k}}$.
After relating
the $\gamma$'s for particle and anti-particle and carrying out  the usual
identification
between the angles in the bootstrap equation and the integers $\xi(t)$, we
derive
$$
\xi(i) - \xi(a) = \xi(b) - \tilde{\xi}(i) + \frac{c(i) }{2} - \frac{ c(a) +
c(b)}{4} \;\;\;   .
$$
Since both sides of this equation have to be integers we can deduce that the
colours of
$a$ and $b$ have to be the same, i.e. $ c(a) = c(b) $.  By eliminating particle
$k$ and
$\bar{k}$ the fusing rule related to this process reads
\beq
\left( \sigma^{ \xi(i) - \xi(a) }  + \sigma^{  \xi(a)  - \xi(i)  + \frac{ c(i)
-
c(a)}{2}   } \right) \gamma_j
\; =\; \gamma_a + \gamma_b \;\; .
\eeq
Similar arguments hold for higher order poles and in general we will have
(\ref{eq: Sbox}).
\par
Employing then the explicit expression for $S$ \cite{FO}, parameterising the
fusing angle
by
$\eta_{j0}^{\alpha} = (2l  + \frac{ c(i) - c(j)}{2} ) \frac{\pi}{h} $ and
shifting the dummy variable
of the product appropriately we obtain
$$
\frac{ \left[ 2q - \frac{ c(l) +1}{2} \right]^{ - \frac{1}{2}  \lambda_l \cdot
\left( \sigma^{q-l +
\frac{ c(j) -c(i)}{2} }+ \sigma^{q+l} \right) \gamma_{j} } }
{ \left[ - 2q + \frac{ c(l) +1}{2} \right]^{ - \frac{1}{2}  \lambda_l \cdot
\left( \sigma^{q+l +
\frac{ c(i) - 1}{2} }+ \sigma^{q-l + \frac{ c(j) - 1}{2}   } \right) \gamma_{j}
} }  =
\frac{ \left[ 2q - \frac{ c(l) +1}{2} \right]^{ - \frac{1}{2}  \lambda_l \cdot
\sigma^{q}
\left( \gamma_{i_{1}} + \ldots  + \gamma_{i_{n} }  \right)     } }
{ \left[ - 2q + \frac{ c(l) +1}{2} \right]^{ - \frac{1}{2}  \lambda_l \cdot
\sigma^{q + \frac{ c(i) -1}{2} }
\left( \gamma_{i_{1}} + \ldots  + \gamma_{i_{n} }   \right)} }
$$
from which we infer the general fusing rule
\beq
\left( \sigma^l + \sigma^{-l + \frac{ c(j) - c(i)}{2}}  \right) \gamma_i  \; =
\;
\gamma_{i_{1}} + \ldots  + \gamma_{i_{n}} \;\; . \label{eq: fuseg}
\eeq
Implicitly we required here that all the colours on the right hand side
coincide, in other words
there is no fusing rule if the colours differ. Notice further that acting on
this equation with
$\sigma_{-}$ or $\sigma_{+}$, which for the standard fusing rule yields a
second
solution
\cite{FO}, will simply turn this equation into itself and hence this equation
possesses only
one solution. The reason for this will be apparent below.  We may now employ a
relation
found in \cite{FO}
\footnote{The minus sign on the right hand side was missing therein.}
\beq
q(1) \cdot \sigma^l \gamma_i  = - i y_i \sin\left(  \frac{ \pi}{h} \right) e^{
-i
\frac{\pi}{h} \left(
2l + \frac{ ( 1- c(i) )}{2}  \right) }
\eeq
and project (\ref{eq: fuseg}) into the velocity plane, then
\beq
2 y_j \cos\left( \eta_{j0}^{\alpha} \right) \; = \;  y_{i_{1}} + \ldots  +
y_{i_{n}} \;\;   .
\eeq
Using now the proportionality between the Perron-Frobenius vector and the
masses
\cite{FLO} yields for the binding energy of level $\alpha$, after setting $E_0
=0$,
\beq
E_{\alpha} \; =\; \frac{1}{2} \left(  m_{i_{1}} + \ldots  + m_{i_{n}} \right)
\;
=\; m_j
\cos\left( \eta_{j0}^{\alpha} \right) \;\;  .
\eeq
We illustrate this in figure 4.
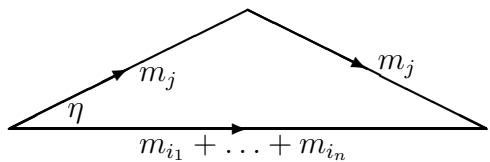
\begin{figure}[h]
\setlength{\unitlength}{0.0125in}
\begin{picture}(40,120)(60,470)
\thicklines
\put(200,510){\line(1,0){200}}
\put(300,560){\line(2,-1){100}}
\put(300,560){\line(-2,-1){100}}
\put(200,510){\vector(1,0){100}}
\put(200,510){\vector(2,1){50}}
\put(300,560){\vector(2,-1){50}}
\put(250,500){  $ m_{i_{1}}  + \ldots  + m_{i_{n}} $ }
\put(250,530){  $ m_j  $ }
\put(350,535){  $ m_j  $ }
\put(220,515){  $ \eta $ }
\end{picture}
 \caption{Mass triangle in the velocity plane}
\end{figure}
 Since this mass-triangle is equilateral  it
becomes clear
as well that there is only one possible triangle, unlike in the general case,
in
which one
can draw an equivalent one, corresponding to the second solution. This is the
geometrical
reason for the fact that the fusing rule possess only one solution.

\section{$E_6$-Affine Toda Field Theory}
\reset

In order to verify the general claims made in the previous section, we shall
investigate now
explicitly $E_6$-affine Toda field theory. It  furnishes a very good example
since it
possesses  a very rich structure and several nontrival features in comparision
with other
cases. Its fusing
properties are
nontrivial, it possesses self-conjugate and non-self-conjugate scalar fields
unlike the other members of the E-series and
in addition, when folded  according to (\ref{eq: symm}), it constitutes the
theory
with the  automorphism of the highest  order, namely 3 .  Its
classical Lagrangian density for the bulk theory reads
\beq
{\cal L}\,=\,\frac{1}{2}\partial_{\mu}\phi\partial^{\mu} \phi -
\frac{m^2}{ \beta ^2}     \sum_{i=0}^{6} n_i e^{  \beta \alpha_i   \cdot \phi}
\eeq
with  $n_i$ denoting the Kac labels, being the integer coefficients in the
expansion of the
highest root in terms of  the simple roots   $\alpha_i$ of $E_6$. $\phi$
constitutes a vector
whose components are the six scalar fields of the theory.
A suitable basis for the simple roots in which the mass-matrix diagonalises is
given by
\beq
{\cal A} = \left( \begin{array}{llllll}
\frac{1}{2} &   -\frac{1}{2}    & -\frac{1}{2} &   -\frac{1}{2}& -\frac{1}{2}
  & -\frac{\sqrt{3}}{2} \\
   1     & 1 &    0     & 0   & 0   & 0 \\
  -1     & 1 &    0     & 0   & 0   & 0 \\
0     & -1 &    1     & 0   & 0   & 0 \\
0     & 0 &    -1     & 1   & 0   & 0 \\
0     & 0 &    0     &  -1   & 1   & 0
\end{array} \right)  \;\;  .
\eeq
Here the entry ${\cal A} _{ij}$ denotes the $j$'s component of the simple root
$\alpha_i$. Labelling
our particles according to the following Dynkin diagram

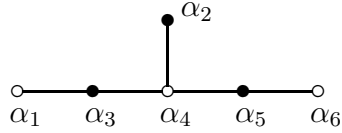
\setlength{\unitlength}{0.01cm}
\begin{picture}( 1000,300)(0,100)
\thicklines
\put(492,190){$ \circ$}
\put(510,200){\line( 1, 0){85}}
\put(592,190){$ \bullet$}
\put(702,207){\line( 0, 1){85}}
\put(692,285){$ \bullet$}
\put(610,200){\line( 1, 0){85}}
\put(692,190){$ \circ$}
\put(792,190){$ \bullet$}
\put(892,190){$ \circ$}
\put(710,200){\line( 1, 0){85}}
\put(810,200){\line( 1, 0){85}}
\put(492,160){$ \alpha_1$}
\put(592,160){$ \alpha_3$}
\put(692,160){$ \alpha_4$}
\put(792,160){$ \alpha_5$}
\put(892,160){$ \alpha_6$}
\put(720,300){$ \alpha_2$}
\end{picture}
\par
The negative of the highest root  acquires the form $\alpha_0 = - \alpha_1 -
2\alpha_2
- 2\alpha_3 - 3\alpha_4 - 2\alpha_5 - \alpha_6 $.  The theory remains invariant
under
the following transformations of the extended Dynkin diagram
\beq
\alpha_1 \rightarrow \alpha_6 \rightarrow \alpha_0   \qquad ,  \qquad
\alpha_2 \rightarrow \alpha_3 \rightarrow \alpha_5    \qquad  ,  \qquad
\alpha_4 \rightarrow \alpha_4 \label{eq: symm}
\eeq
which may be used to derive the twisted theory $D_4^{(3)}$-theory via the
folding procedure
[27-30].  The masses which in all Toda theories may be organised
into the
Perron-Frobenius eigenvector of the Cartan matrix take on particularly simple
values
\bea
m_1^2  &=&  m_6^2 = ( 3 - \sqrt{3}) \; m^2    \qquad \qquad  m_2^2 =\, 2 (3 -
\sqrt{3} ) \; m^2 \\
m_3^2  &=&  m_5^2 = ( 3 + \sqrt{3}) \; m^2   \qquad \qquad  m_4^2 =  2 (3 +
\sqrt{3} ) \; m^2 \;\;  .
\eea
For reasons of completeness we shall furthermore report all the two-particle
scattering matrices up to
the ones which can be trivially  obtained  from the identities $S_{ij}(\th) =
S_{ji}(\th) = S_{\ib \jb}(\th)$
\cite{BCDS}
$$
 \begin{array}{llll}
S_{11}  = \{1\} \{7\} & S_{12}  = \{4\} \{8\}     & S_{13} = \{2\} \{6\} \{8\}
 \\
S_{14}  = \{3\} \{5\} \{7\} \{9\} & S_{15}  = \{4\} \{6\}\{10\}     & S_{16}  =
\{5\} \{11\}    \\
S_{22} = \{1\} \{5\} \{7\} \{11\} & S_{23}  = \{3\} \{5\}\{7\} \{9\}    &
S_{24}
 = \{2\} \{4\} \{6\}^2 \{8\}
 \{10\}    \\
S_{33} = \{1\} \{3\} \{5\} \{7\}^2 \{9\}    & S_{34} = \{2\} \{4\}^2 \{6\}^2
\{8\}^2 \{10\}    &
S_{35}  = \{3\} \{5\}^2 \{7\} \{9\} \{11\}
\end{array}
$$
$$ S_{44}  = \{1\} \{3\}^2 \{5\}^3 \{7\}^3 \{9\}^2   \{11\} \;\; . $$
We have omitted here the explicit dependence on the rapidity $\th$.
Then by starting with one of the solutions obtained for
the boundary in the ground
state \cite{FK1}
\bea
W_{10}(\th)  &=&  W_{60}(\th)   = {\cal W}_5 (\th){\cal W}_{35}(\th) \\
W_{20}(\th) &=& {\cal W}_1(\th) {\cal W}_{7}(\th)  {\cal W}_{11}(\th) {\cal
W}_{29}(\th)\\
W_{30}(\th) &=& W_{50}(\th) = {\cal W}_3(\th) {\cal W}_{5}(\th)  {\cal
W}_{7}(\th)  {\cal W}_{11}(\th)
                           {\cal W}_{29}(\th) {\cal W}_{33}(\th)\\
W_{40}(\th) &=& {\cal W}_1(\th) {\cal W}_{3}(\th)  {\cal W}_{5}^2(\th)  {\cal
W}_{7}(\th)
                           {\cal W}_{9}^2(\th) {\cal W}_{27}(\th)  {\cal
W}_{29}(\th){\cal W}_{31}^2(\th)
                       {\cal W}_{35}(\th)
\eea
we now construct all the higher boundary states by means of a bootstrap
principle which is
in close analogy to the one in the bulk theory. Employing  the fact that the
poles of  ${\cal W}_x (\th)$
are situated at $ \th_{\pm} = \frac{ \pm 1 - x -h}{2h} i \pi $, with  $h=12$
and $\pm$
refering to the sign of the residues,
we observe that the only possible poles in the physical sheet are the ones
resulting from the blocks
which have been shifted. Letting then the odd order poles with positive residue
participate in the bootstrap
equation (\ref{eq: bbsb}) we compute
\bea
W_{j \alpha} (\th) &=& S_{j 1}\left( \th - \frac{i \pi}{12} \right) W_{j  0}
(\th)  S_{j 1}\left( \th  + \frac{i \pi}{12} \right) \; =\;  S_{j 3}\left( \th
\right)
W_{j  0} (\th)     \label{eq: Walpha}  \\
W_{j \beta} (\th) &=& S_{j 2}\left( \th - \frac{4 i \pi}{12} \right) W_{j  0}
(\th)  S_{j 2}\left( \th + \frac{4 i \pi}{12} \right) \; =\;  S_{j 2}\left( \th
\right)
W_{j  0} (\th)  \\
W_{j \gamma} (\th) &=& S_{j 3}\left( \th - \frac{2i \pi}{12} \right)  W_{j  0}
(\th)  S_{j 3}\left( \th + \frac{2i \pi}{12} \right)  \; =\;  S_{j 2}\left( \th
\right)
 S_{j 5}\left( \th  \right) W_{j  0} (\th)  \\
W_{j \delta} (\th) &=& S_{j 3}\left( \th - \frac{4i \pi}{12} \right) W_{j  0}
(\th)  S_{j 3}\left( \th + \frac{4i \pi}{12} \right)\; =\;  S_{j 5}\left( \th
\right)
 W_{j  0} (\th)  \\
W_{j \epsilon} (\th) &=&  S_{j 4}\left( \th -\frac{i \pi}{12} \right)   W_{j
0} (\th)   S_{j 4}\left( \th +\frac{i \pi}{12} \right) =  S_{j 2}\left( \th
\right)
S_{j 3}\left( \th  \right) S_{j 5}\left( \th  \right) W_{j  0} (\th)
\;\;\;\;\;
\;\; \\
W_{j \phi} (\th) &=& S_{j 4}\left( \th  - \frac{3 i \pi}{12} \right) W_{j  0}
(\th)  S_{j 4}\left( \th + \frac{3 i \pi}{12} \right)\; =\;  S_{j 3}\left( \th
\right)
S_{j 5}\left( \th  \right) W_{j  0} (\th)  \\
W_{j \rho} (\th) &=& S_{j 5}\left( \th \!-\! \frac{2 i \pi}{12} \right)  W_{j
0}
(\th)  S_{j 5}\left( \th \!+ \! \frac{2 i \pi}{12} \right) \; =\;  S_{j
2}\left(
\th  \right)
S_{j 3}\left( \th  \right) W_{j  0} (\th)  \;\;  .  \label{eq: Wrho}
\eea
Clearly in order to have a proper bootstrap we have to ensure that our system
closes under
equation  (\ref{eq: bbsb})  and we still have to investigate the singularity
structure of
(\ref{eq: Walpha}) -  (\ref{eq: Wrho}). Keeping the same philosophy, that is
bootstraping
only on odd order poles with positive residues we can account for
\underline{all} poles in this manner.
It is worth noting that in proceeding in this fashion care has to be taken
about
possible zeros emerging
from the unshifted blocks which might cancel some of the poles or alter  their
order. In the
following table we present all the ``fusing angles" which participate in the
bootstrap:
\par
\begin{table} [h]
\begin{tabular}{ c | c | c| c | c | c | c| c | c | }
 $ i \setminus \mu $ & 0  & $\alpha$  & $\beta$  &  $\gamma$  &  $\delta$  &
 $\epsilon$  &  $\phi$   &  $\rho$    \\ \hline
 1   &  $1^{\alpha} $                                   & $3^{\gamma}   $
  & $1^{\rho}5^{\alpha} $     & $1^{\epsilon}5^{\phi}11^{\beta}$  &
$1^{\phi}7^{\beta}11^0$  &
   $ 11^{\rho}$      & $7^{\rho} 11^{\alpha}$        & $9^{\delta}$  \\ \hline
 2   &  $4^{\beta}$       &  $4^{\rho} 6^{\alpha}  $  &
  $2^{\phi}6^{\beta}8^0$    & $6^{\gamma}8^{\delta}$   &
$4^{\gamma}6^{\delta}$
     &
    $6^{\epsilon} 8^{\phi}$ &  $4^{\epsilon} 6^{\phi} 10^{\beta}$  & $
6^{\rho}
8^{\alpha}$ \\ \hline
 3   & $ 2^{\gamma} 4^{\delta}   $            &  $ 2^{\epsilon}4^{\phi}8^0   $
&
$ 4^{\gamma}   $  &
      $6^{\rho}$      & $6^{\alpha}$    &  $8^{\gamma} 10^{\delta}$   & $
8^{\delta}$  &
      $4^{\epsilon}  8^{\beta} 10^0$ \\ \hline
 4   &  $1^{\epsilon}3^{\phi}5^{\beta}  $ & $ 5^{\rho}  $ & $ 3^{\epsilon}7^0 $
  &
      $7^{\delta}$  &  $5^{\gamma}$      &  $ 7^{\phi} 9^{\beta} 11^0$
    & $5^{\epsilon} 9^0$         & $ 7^{\alpha}$  \\ \hline
 5   &  $ 2^{\rho}4^{\alpha} $                    &  $6^{\delta}    $     &  $
4^{\rho}$     &
     $4^{\epsilon}8^{\beta}10^0$& $2^{\epsilon}4^{\phi}8^0$      &$8^{\rho}
10^{\alpha} $
       & $ 8^{\alpha}$    &   $ 6^{\gamma}$ \\ \hline
 6   &   $1^{\delta}  $                                   &
$1^{\phi}7^{\beta}11^0  $ & $ 1^{\gamma} 5^{\delta}$
     &    $9^{\alpha}$    & $ 3^{\rho}$ & $11^{\gamma}$        & $7^{\gamma}
11^{\delta}$
             &  $1^{\epsilon} 5^{\phi} 11^{\beta}$   \\ \hline
\end{tabular}
\caption{The boundary fusing angles $\eta_{i \mu}^{\nu}$}
\end{table}
\par
Here each entry in the table indicates a fusing angle as multiple of $\frac{i
\pi}{12}$, where the left column
refers to the particle type which scatters off the boundary in the state
indicated in the first row. The
superscript refers to the state the boundary is changing into.
We  observe that all the angles match up in the way announced earlier in
equation
(\ref{eq: zwei}). As a  further consistency  check we may
then employ  the  equation  subsequent to  (\ref{eq: zwei})  together with the
explicit form for the scattering matrices.
In figure 5 we illustrate for some cases which particles may cause an
excitation
or
lowering of the states  in the boundary when
possessing a particular value of the rapidity. So we obtain an interesting
picture familiar from
atomic physics suggesting that the boundary interacts with the particles in a
kind  of matter-radiation way. Due to the CDD-ambiguity, there is a second
solution
for the ground state and a similar result may be found, when starting with
 this other  ground state.
The ambiguity mentioned at the end of section 2 has now  disappeared. We may
start
the bootstrap with an expression $ \prod S W_{i0}$, where the product runs over
a different range than the one found in our solution.  For the cases we checked
we do
not obtain a closed bootstrap. However, there may be possibilities for which
this can
be achieved. Appealing to a similar principle as in the bulk theory, that is
expecting the
solution to be ``minimal", the presented solution seems the
 most natural one.

\section{Conclusions}
\reset
As our principal result we have found that the generalization of the
Coleman-Thun mechanism
in the bulk theory possesses an analogue in the theory with boundaries. For the
$E_6$-affine
Toda field theory we have explicitly demonstrated how the bootstrap closes. The
binding
energies are half the sum of the masses belonging to the same colour with
respect to the
bicolouration of the Dynkin diagram. We found six different energy levels in
this case, of
which two are degenerate.
\par
We have carried out a similar analysis for the $E_7$-theory and several members
of the
A-series, finding that the bootstrap always closes in this fashion. We obtain
in
each case
that the number of energy levels equals the rank of the
 underlying Lie algebra.
\par
There are several immediate questions to be answered. The generalisation of
this result to
other theories is highly desirable. Having a closed formula for the  W-matrix
in
analogy to
the S-matrix, valid for all algebras will probably be a necessary step into
this
direction.
A complete investigation of the possible integrable boundary conditions is
still
outstanding,
although steps into this direction have been undertaken in \cite{CDRS}.
\par
A very interesting challenge is posed by the  question of how to extend this
results to the
situation off-shell.
\par
\begin{figure}[h]
\setlength{\unitlength}{0.0125in}
\begin{picture}(40,400)(30,470)
\thicklines
\put(100,500){\line(1,0){420}}
\put(100,580){\line(1,0){420}}
\put(100,797){\line(1,0){420}}
\put(100,608){\line(1,0){420}}
\put(100,688){\line(1,0){420}}
\put(100,717){\line(1,0){420}}
\put(100,500){\vector(0,1){350}}
\put(80,840){ $ E$}
\put(50,580){ $ \frac{1}{2} m_2$}
\put(-3,797){ $ \frac{1}{2} (m_2 + m_3 + m_5)$}
\put(50,608){ $ \frac{1}{2} m_3 $}
\put(27,688){ $ \frac{1}{2} (m_2 + m_5) $}
\put(27,717){ $ \frac{1}{2} (m_3 + m_5)$}
\put(521,498){ $  0 $}
\put(521,578){ $  \beta $}
\put(521,606){ $  \alpha,\delta $}
\put(521,686){ $  \gamma,\rho $}
\put(521,715){ $  \phi $}
\put(521,795){ $  \epsilon $}
\put(80,500){ $  0 $}
\thinlines
\put(110,500){\vector(0,1){80}}
\put(111,505){ $ 4 $}
\put(111,568){ $ 4 $}
\put(160,500){\vector(0,1){80}}
\put(161,505){ $ 2 $}
\put(161,568){ $ 2 $}
\put(210,500){\vector(0,1){108}}
\put(211,505){ $ 6 $}
\put(211,596){ $ 1 $}
\put(260,500){\vector(0,1){108}}
\put(261,505){ $ 3 $}
\put(261,596){ $ 5 $}
\put(310,500){\vector(0,1){188}}
\put(311,505){ $ 5 $}
\put(311,676){ $ 3 $}
\put(360,500){\vector(0,1){217}}
\put(361,505){ $ 4 $}
\put(361,705){ $ 4 $}
\put(410,500){\vector(0,1){297}}
\put(411,505){ $ 4 $}
\put(411,785){ $ 4 $}
\put(460,500){\vector(0,1){188}}
\put(461,505){ $ 3 $}
\put(461,676){ $ 5 $}
\put(510,500){\vector(0,1){108}}
\put(495,505){ $ 5 $}
\put(495,596){ $ 3 $}
\put(110,510){\vector(0,-1){10}}
\put(160,510){\vector(0,-1){10}}
\put(210,510){\vector(0,-1){10}}
\put(260,510){\vector(0,-1){10}}
\put(310,510){\vector(0,-1){10}}
\put(360,510){\vector(0,-1){10}}
\put(410,510){\vector(0,-1){10}}
\put(460,510){\vector(0,-1){10}}
\put(510,510){\vector(0,-1){10}}
\put(135,580){\vector(0,1){28}}
\put(135,590){\vector(0,-1){10}}
\put(136,585){ $ 6 $}
\put(136,596){ $ 1 $}
\put(110,608){\vector(0,1){80}}
\put(110,618){\vector(0,-1){10}}
\put(111,613){ $ 6 $}
\put(111,676){ $ 1 $}
\put(160,608){\vector(0,1){80}}
\put(160,618){\vector(0,-1){10}}
\put(161,613){ $ 2 $}
\put(161,676){ $ 2 $}
\put(235,608){\vector(0,1){109}}
\put(235,618){\vector(0,-1){10}}
\put(236,613){ $ 5 $}
\put(236,705){ $ 3 $}
\put(285,608){\vector(0,1){189}}
\put(285,618){\vector(0,-1){10}}
\put(286,613){ $ 5 $}
\put(286,780){ $ 3 $}
\put(135,688){\vector(0,1){29}}
\put(135,698){\vector(0,-1){10}}
\put(136,693){ $ 6 $}
\put(136,705){ $ 1 $}
\put(210,688){\vector(0,1){109}}
\put(210,698){\vector(0,-1){10}}
\put(211,693){ $ 6 $}
\put(211,785){ $ 1 $}
\put(510,688){\vector(0,1){109}}
\put(510,698){\vector(0,-1){10}}
\put(495,693){ $ 5 $}
\put(495,785){ $ 3 $}
\put(110,717){\vector(0,1){80}}
\put(110,727){\vector(0,-1){10}}
\put(111,722){ $ 2 $}
\put(111,785){ $ 2 $}
\put(160,717){\vector(0,1){80}}
\put(160,727){\vector(0,-1){10}}
\put(161,722){ $ 4 $}
\put(161,785){ $ 4 $}
\end{picture}
 \caption{Boundary transitions for $E_6$ affine Toda field theory}
\end{figure}
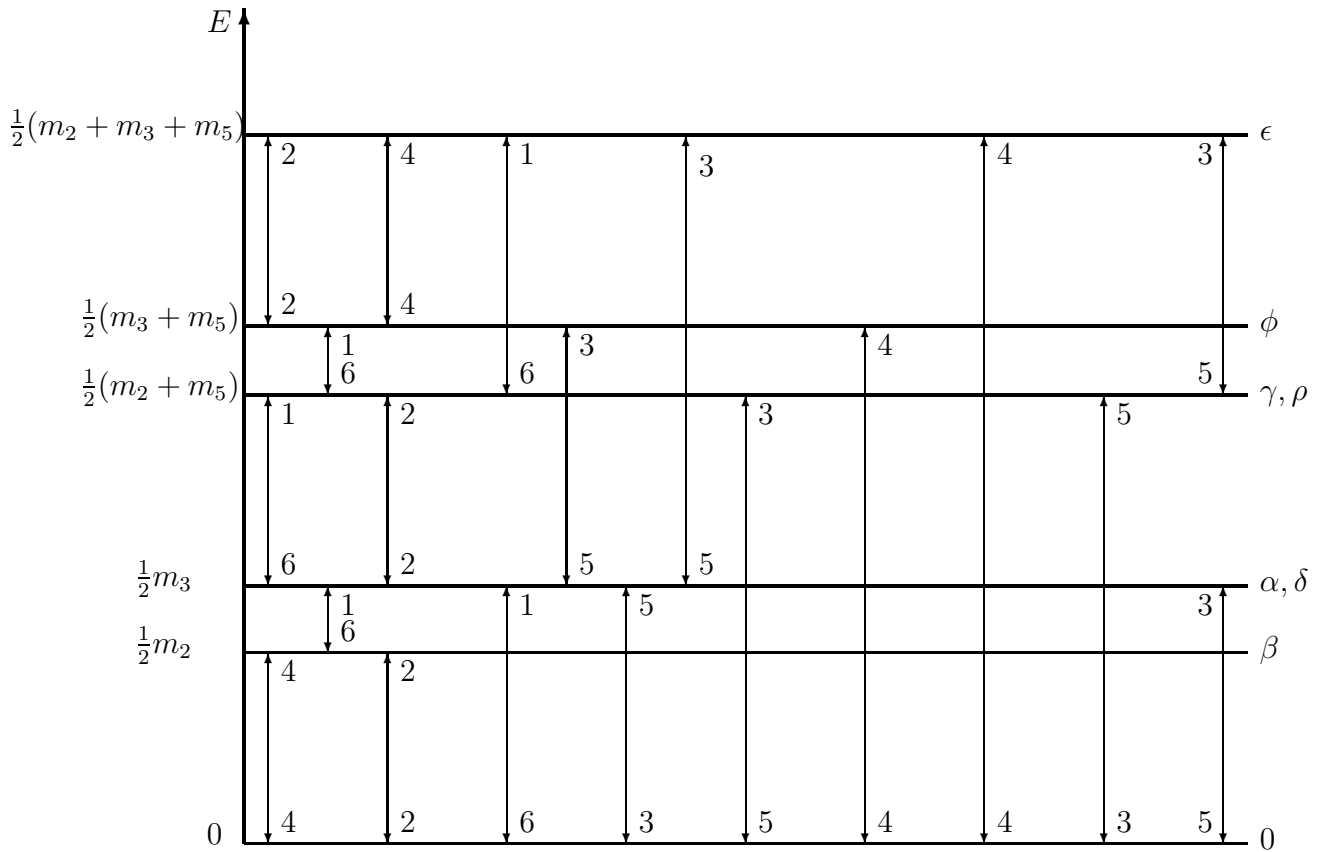

{\bf Acknowledgements}

A.F. would like to thank HEFCW and FAPESP(Brasil) for financial support and
furthermore the
Instituto de Fisica de S\~ao Carlos for its hospitality, where part of this
work
has been carried
out.

After the completion of our work we received a recent preprint \cite{CDRS}, in
which similar
conclusions have been reached for the case of $A_2$-affine Toda field theory.


\begin{thebibliography}{99}
\bibitem{YB}  C.N. Yang, {\em Phys. Rev. Lett.} {\bf 19} (1967) 1312;
              R.J. Baxter, {\em Exactly Solved Models
              in Statistical Mechanics} (Academic Press, London, 1982).
\bibitem{ZZ} A.B. Zamolodchikov and Al. B. Zamolodchikov, {\em Ann. Phys.}
{\bf 120} (1979) 253.
\bibitem{Leg} A.O. Caldeira and A.J. Leggett, {\em Phys. Rev. Lett.}
 {\bf 46} (1981) 211; {\em Physica} {\bf A121} (1983) 587; {\em Ann. Phys.}
{\bf 149} (1983) 374.
\bibitem{Callan} C.G. Callan and L. Thorlacius, {\em Nucl. Phys. } {\bf B329}
(1990) 117.
\bibitem{Witten} C.G. Callan, C. Lovelace, C.R. Nappi and S.A. Yost,
{\em Nucl. Phys.} {\bf B293} (1987) 83; E.Witten, {\em Phys. Rev.} {\bf D46}
 (1992) 5467; K. Li and E. Witten, Princeton-preprint IASSNS-HEP-93/7.
\bibitem{Ch1} I.V. Cherednik {\em Theor. and Math. Phys.} {\bf 61} {1984}
{977}.
\bibitem{FK} A. Fring and R. K\"oberle, {\em Factorized Scattering in the
Presence of Reflecting Boundaries},  S\~ao Carlos-preprint USP-IFQSC/TH/93-06,
hep-th/9304141,\NP {\bf B} in print.
\bibitem{GZ} S. Ghoshal and A. Zamolodchikov, {\em Boundary S-matrix and
Boundary
State in Two-Dimensional Integrable Quantum Field theory}, Rutgers-preprint
RU-93-20, hep-th/9306002.
 \bibitem{FK1} A. Fring and R. K\"oberle, {\em  Affine Toda Field Theory in the
 Presence of Reflecting Boundaries} ,  S\~ao Carlos preprint
  USP-IFQSC/TH/93-12, hep-th/9309142,  \NP {\bf B} in print.
\bibitem{Ch2} I.V. Cherednik, {\em Notes on affine Hecke algebras. 1.
Degenerated affine Hecke algebras and Yangians in mathematical physics.},
BONN-HE-90-04 .
\bibitem{Sk} E.K. Sklyanin, {\em J. Math. Phys.} {\bf A21} (1988) 2375.
\bibitem{MN} L. Mezincescu and R.I. Nepomechie, {\em J. Phys. A: Math. Gen.}
 {\bf 25} (1992) 2533; L. Mezincescu, R.I. Nepomechie and V. Rittenberg,
 {\em  Phys. Lett.} {\bf A147} (1990) 70;
L. Mezincescu and R.I. Nepomechie, {\em Argonne Workshop on Quantum Groups}
ed. T. Curtright, D. Fairlie and C. Zachos (World Scientific, Singapore, 1991);
L. Mezincescu and R.I. Nepomechie, {\em Quantum Field Theory, Statistical
Mechanics, Quantum Groups and Topology} ed. T. Curtright, L. Mezincescu and
R.I.
Nepomechie (World Scientific, Singapore, 1992).
\bibitem{KSA} P.P. Kulish and R. Sasaki, {\em Prog. Theor. Phys.} {\bf 89}
(1993) no. 3;
P.P. Kulish, R. Sasaki and C. Schwiebert, {\em J. Math. Phys.} {\bf 34} (1993)
 286.
\bibitem{DeVega} H.J. De Vega and A. Gonz\'alez-Ruiz, {\em Boundary
K matrices for the six vertex and the $n(2n-1) A_{n-1}$ vertex models}
Paris-preprint LPTHE-PAR-92-45; {\em Boundary K matrices
 for the XYZ, XXZ and XXX spin chains } Paris-preprint LPTHE-PAR-93-29.
\bibitem{Ghoshal} S. Ghoshal, {\em Bound State Boundary S-matrix of the
Sine-Gordon Model},  Rutgers preprint RU-93-51, hep-th/9310188; {\em
Boundary S-Matrix  of the O(n) Symmetric Nonlinear Sigma Model},
 Rutgers preprint RU-94-02, hep-th/9401008.
\bibitem{Sasaki} R. Sasaki,  {\em Reflection Bootstrap Equations for Toda Field
Theory},  Kyoto-preprint   YITP/U-93-33, hep-th/93110027.
\bibitem{Chim} L. Chim, {\em Boundary S-matrix for the Integrable q-Potts
Model},
 Rutgers preprint RU-94-33, hep-th/9404118.
\bibitem{FS} P. Fendley  and H. Saleur, {\em Deriving boundary S matrices},
 preprint USC-94-001, hep-th/9402045.
\bibitem{MOP} A.V. Mikhailov, M.A. Olshanetsky and A.M. Perelomov,
{\em Comm. Math. Phys.} {\bf 79} (1981), 473; G. Wilson, {\em Ergod. Th.
Dyn. Syst.} {\bf 1} (1981) 361; D.I. Olive and  N. Turok, {\em Nucl. Phys.}
{\bf B257} [FS14] (1985) 277.
\bibitem{Muss} P. Christe and G.Mussardo, {\em Nucl. Phys.} {\bf B330} (1990)
465, {\em Int. J. of Mod. Phys.} {\bf A5} (1990) 4581.
\bibitem{BCDS} H. W. Braden, E. Corrigan, P. E. Dorey and R. Sasaki, {\em
Phys. Lett.} {\bf B227} (1989) 411 and {\em Nucl. Phys.} {\bf B338} (1990) 689.
\bibitem{Coleman} S. Coleman and H. Thun, {\em Commun. Math. Phys.} {\bf 61}
(1978) 31.
\bibitem{CDD} L. Castillejo, R.H. Dalitz and F.J. Dyson, {\em Phys. Rev.}
{\bf 101} (1956), 453.
\bibitem{PD} P.E. Dorey, {\em Nucl. Phys.} {\bf B358} (1991) 654;
 {\em Nucl. Phys.} {\bf B374} (1992) 741.
\bibitem{FLO} A. Fring, H.C. Liao and D.I. Olive, {\em Phys. Lett.} {\bf B266}
(1991) 82.
\bibitem{FO} A. Fring and D.I. Olive, {\em Nucl. Phys.} {\bf B379} (1992) 429.
\bibitem{SH} S. Helgason, {\em Differential Geometry and Symmetric Spaces}
(Academic Press, London, 1978).
\bibitem{OT} D.I. Olive and  N. Turok, {\em Nucl. Phys.} {\bf B215} [FS7]
(1983) 470.
\bibitem{FK2} A. Fring and R. K\"oberle, {\em On exact S-matrices for
non-simply laced affine Toda theories},  S\~ao Carlos-preprint
USP-IFQSC/TH/93-13.
\bibitem{KO}  M.A.C. Kneipp and  D.I. Olive,  {\em  Solitons and Vertex
Operators in Twisted
Affine Toda Field Theories},  Swansea preprint  SWAT/93-94/19, hep-th/9304030.
\bibitem{CDRS} E. Corrigan, P.E. Dorey, R.H. Rietdijk and R. Sasaki, {\em
Affine Toda field theory on a half-line} Durham/Kyoto-preprint DTP-94/7,
YITP/U-94-11,
hep-th/9404108.
\end{thebibliography}
\end{document}